\documentclass[twocolumn,showpacs,aps,prl,superscriptaddress,floatfix]{revtex4}
\usepackage{epsfig}
\usepackage{xy}
\begin{document}

\title{ Measurement of the damping of nuclear shell effect  in the doubly
magic $^{208}$Pb region} 

\author{P. C. Rout}
\thanks{email: prout@barc.gov.in}
\affiliation{Nuclear Physics Division, Bhabha Atomic Research Centre, Mumbai 400085, India}
\affiliation{Homi Bhabha National Institute, Anushaktinagar, Mumbai 400094, India}
\author{D. R. Chakrabarty}
\affiliation{Nuclear Physics Division, Bhabha Atomic Research Centre, Mumbai 400085, India}
\affiliation{Homi Bhabha National Institute, Anushaktinagar, Mumbai 400094, India}
\author{V. M. Datar}
\affiliation{Nuclear Physics Division, Bhabha Atomic Research Centre, Mumbai 400085, India}
\affiliation{Homi Bhabha National Institute, Anushaktinagar, Mumbai 400094, India}
\author{Suresh Kumar}
\affiliation{Nuclear Physics Division, Bhabha Atomic Research Centre, Mumbai 400085, India}
\affiliation{Homi Bhabha National Institute, Anushaktinagar, Mumbai 400094, India}
\author{E. T Mirgule}
\affiliation{Nuclear Physics Division, Bhabha Atomic Research Centre, Mumbai 400085, India}
\author{A. Mitra}
\affiliation{Nuclear Physics Division, Bhabha Atomic Research Centre, Mumbai 400085, India}
\author{V. Nanal }
\affiliation{Department of Nuclear and Atomic Physics, Tata Institute of Fundamental Research, Mumbai 400005, India}
\author{S. P. Behera}
\affiliation{Nuclear Physics Division, Bhabha Atomic Research Centre, Mumbai 400085, India}
\affiliation{Homi Bhabha National Institute, Anushaktinagar, Mumbai 400094, India}
\author{V. Singh}
\affiliation{Homi Bhabha National Institute, Anushaktinagar, Mumbai 400094, India}
\affiliation{India based Neutrino Observatory, Tata Institute of Fundamental Research, Mumbai 400005, India}

\begin{abstract}
The damping of the nuclear shell effect with excitation energy has been measured through an analysis of 
the neutron spectra following the triton transfer in the $^7$Li induced reaction on $^{205}$Tl.
The measured neutron spectra demonstrate  the expected large shell correction energy
for the nuclei in the vicinity of doubly magic $^{208}$Pb and a small value around $^{184}$W.
A quantitative extraction of the allowed values of the damping parameter $\gamma$, along with those for the asymptotic nuclear level density parameter $\tilde{a}$, has been made for the first time. 

\end{abstract}

\maketitle
The shell effect is a cornerstone of the mean field theory describing finite 
fermionic systems. The shell structure in atoms decides the chemical properties
of the corresponding elements. In nuclear physics the spin orbit coupling, in addition, plays a dominant
role in deciding the shell closures and the associated magic numbers of protons
and neutrons. The nuclei having such numbers of  neutrons and protons have an 
extra stability with respect to that expected from the average behaviour 
described by the liquid drop model (LDM).
Many important nuclear phenomena such as the occurrence of super heavy elements~\cite{z117, hofmann}, fission isomers~\cite{poli, strut}, super-deformed nuclei~\cite{jans} and new magic numbers in exotic nuclei~\cite{hof, kan} are the consequences of the shell effect. 
The shell effect also affects another 
fundamental property of the nucleus viz. the nuclear level density~(NLD).   
The NLD is an indispensable input to the
statistical calculation of compound nuclear decay and thus an important
physical quantity for many practical applications, such as the calculations of reaction rates 
relevant to nuclear astrophysics, nuclear reactors and spallation
neutron sources.

The NLD was first calculated by Bethe using a non-interacting Fermi gas 
model, without shell effects, arriving at its leading dependence on excitation energy~(E$_X$)
 and angular momentum~($J$)\cite{bethe, bohr}.
The generic behaviour with respect to E$_X$ is described by e$^{2\sqrt{aE_X}}$. Here {\it `a'} is 
the NLD parameter which is related to the single particle density at the Fermi energy.
Direct measurements of the NLD are based on the study of slow neutron resonances, which are mainly
s- and p-wave, and are extrapolated to higher $J$ values to estimate the angular momentum summed or total NLD.
The total NLD inferred from such a measurement shows that on the average the level density
parameter $a$ increases linearly with the mass number (A) of the nucleus as $a \approx$A/8~MeV$^{-1}$. However, there is
a significant departure from this liquid drop value at shell closures. This departure is the largest for the doubly magic nucleus
 $^{208}$Pb, where $a$ (at E$_X\sim$7 MeV) is as low as A/26~MeV$^{-1}$. This shell effect 
on the NLD parameter is expected to damp with excitation energy so that $a$ approaches 
its liquid drop value at E$_X\sim$~40 MeV~\cite{vsr}. It is important to make measurements  on the damping of the shell effect over a wide E$_X$ range. To our knowledge, no such measurement has been reported.

Experimental information on the damping of the shell effect can be obtained by measuring the E$_X$ 
dependence of the NLD over a wide range, typically $\sim$~5-40~MeV. One method, which is limited to the particle bound excitation energy region, involves  
the measurement of continuum $\gamma$-ray spectra following inelastic scattering and 
transfer reactions\cite{melby}. Both NLD and 
$\gamma$-ray strength function are inputs to the analysis of the spectra.  
Syed {\it et al.} used $^3$He induced inelastic scattering and single nucleon transfer reaction
to populate $^{205-208}$Pb \cite{syed} and extracted the energy dependence of NLD from the 
coincident $\gamma$ spectrum up to E$_X$ $\sim$~6~MeV. Another method of addressing the E$_X$ dependence
of NLD over a wider range is by measuring particle evaporation spectra following 
heavy ion fusion reaction and using a statistical model analysis~\cite{drc}. 
Lunardon et al.‪\cite{luna} measured proton evaporation spectra in $^{10,11}$B+$^{198}$Pt reactions and  
extracted the NLD in $^{208}$Pb at an excitation energy $\sim$ 50 MeV. However, this 
excitation energy is above the region influenced by the shell effect and  the extracted
NLD showed the expected liquid drop behaviour. 
It is indeed difficult to access a much lower excitation energy region 
using such heavy ion fusion reaction because of the large Coulomb barrier in the entrance channel.
One way out of this difficulty is to measure particle evaporation spectra following transfer induced fusion
process populating particle unbound states. 
\begin{figure}
\begin{xy}
0*{\includegraphics[width=75mm,angle=90]{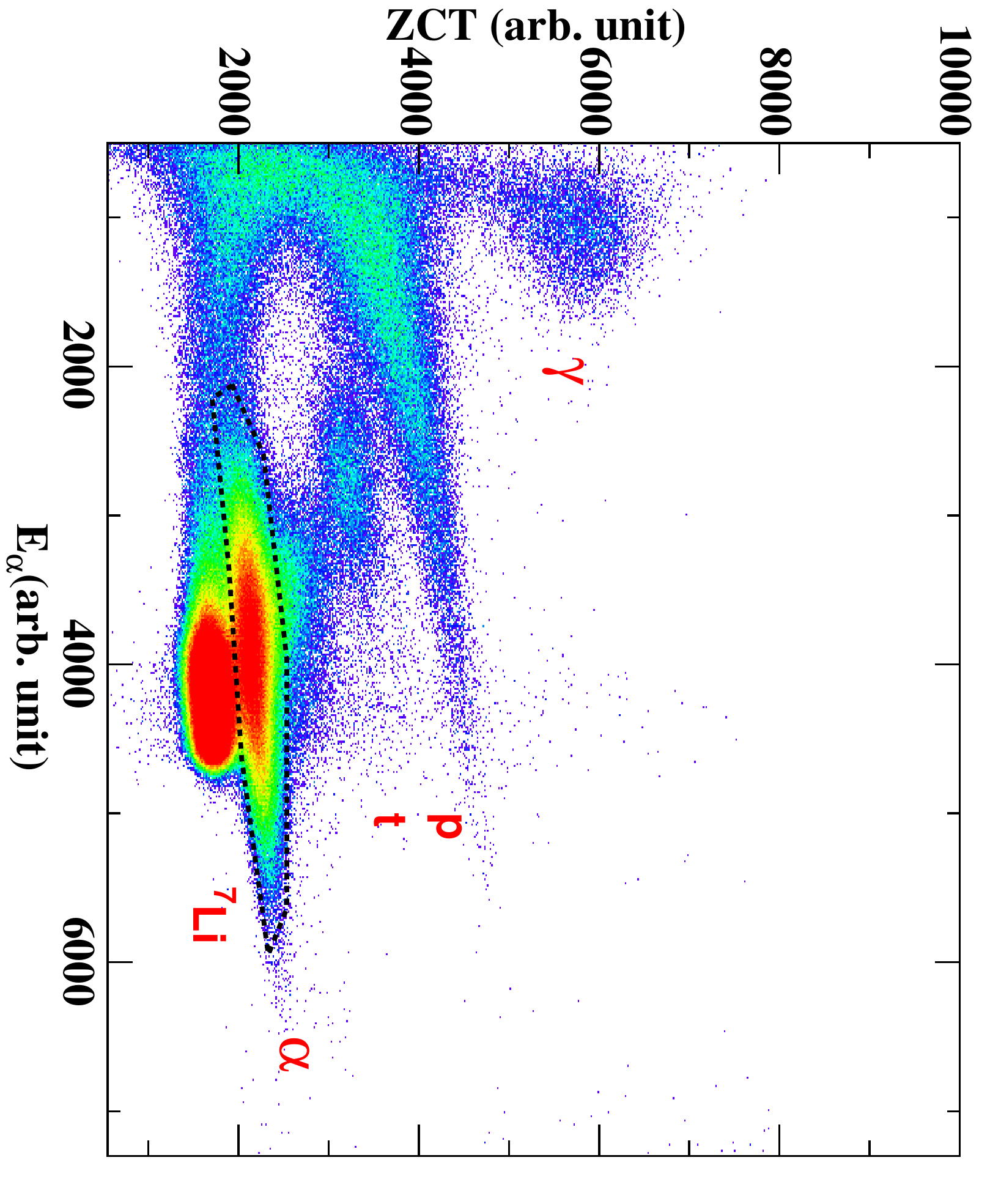}}
+/ur2.5cm/*{\includegraphics[width=34.5mm,height=48.5mm,angle=90]{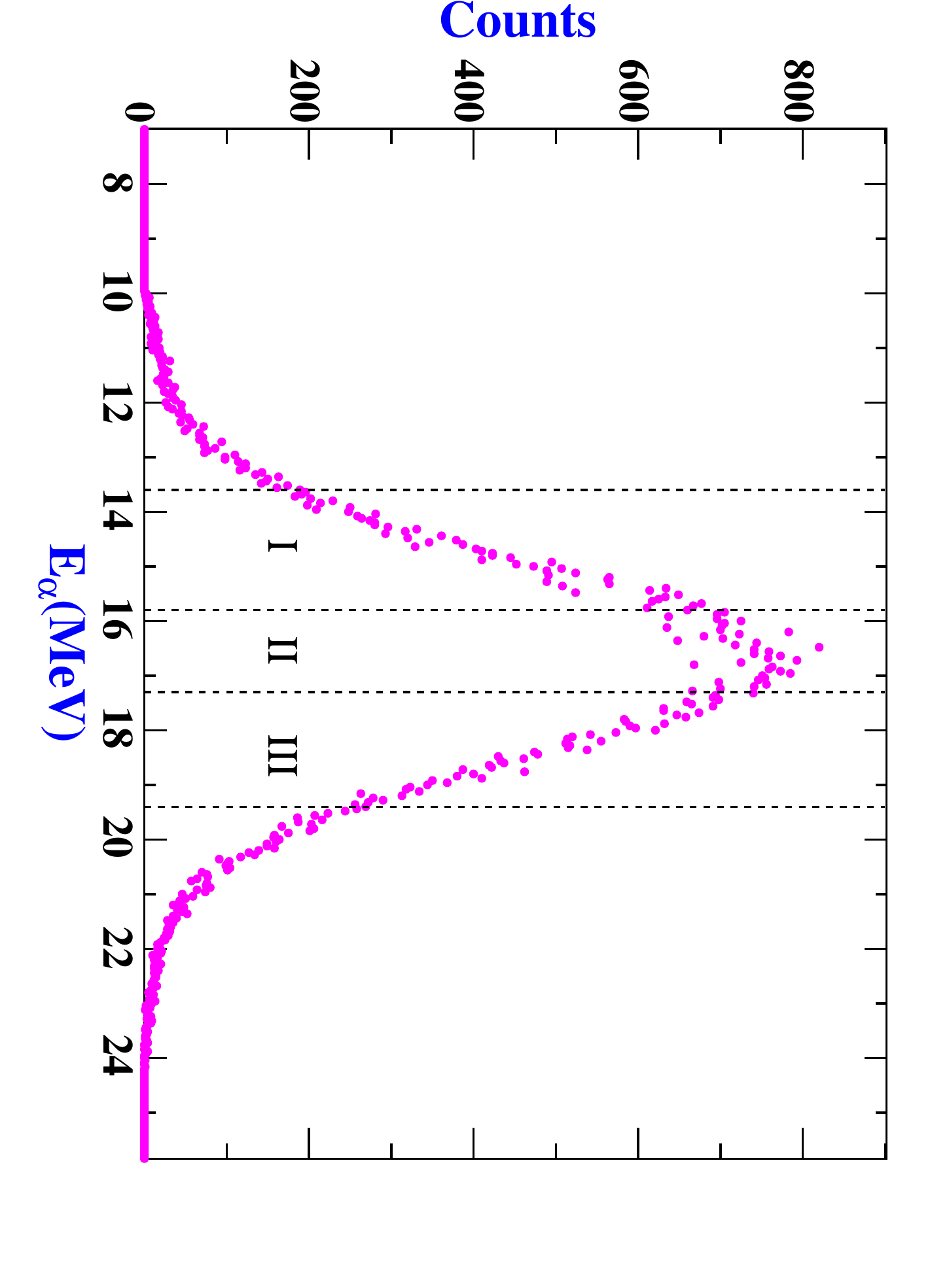}}
\end{xy}
\caption{\label{fig1}(Color online) Two dimensional plot of ZCT vs energy deposited 
in one of the CsI(Tl) detectors in $^7$Li+ $^{205}$Tl reaction.
The inset shows the projected spectrum of the alpha particles defined by the
dotted two dimensional gate. The vertical lines define three alpha energy bins (see text).}
\end{figure}

In this letter, we present an exclusive measurement of neutron spectra from $^{208}$Pb, following triton transfer in the $^{7}$Li+$^{205}$Tl reaction, in coincidence with ejectile alpha particles. The nucleus $^{208}$Pb (formed in the excitation energy range 19~-~23~MeV in this work) decays predominately by first step neutron emission populating $^{207}$Pb in the E$_X\sim$~3~-~14 MeV. Over this  E$_X$ range, the NLD parameter is expected to show a significant change due to the damping of the shell effect.
We have also made the above measurement with a $^{181}$Ta target populating nuclei in the $^{184}$W region where the shell effect is expected to be small.

The experiment was performed at the Mumbai Pelletron Linac Facility
(PLF) using a 30 MeV pulsed $^7$Li beam of width $\sim$1.5 ns (FWHM) and period $\sim$~107~ns. 
Self-supporting foils of 4.7 mg/cm$^2$  $^{205}$Tl (enriched to $>$99\%) and 3.7
mg/cm$^2$ $^{181}$Ta ($\sim$100\% natural abundance) were used as targets. 
Alpha particles were detected at backward angles ($\sim$126$^{\circ}$-150$^{\circ}$)
in 8 CsI(Tl) detectors of dimensions 2.5~cm $\times$ 2.5~cm $\times$ 1~cm (thick) coupled to
Si(PIN) photodiodes and  placed at a distance of $\sim$ 5~cm from the target. The detectors were 
covered with an aluminised mylar foil of thickness $\sim$1~$\mu$m. Particle identification
was done using the standard  pulse shape discrimination method by measuring the zero
cross over timing (ZCT) of the amplified bipolar pulse.

\begin{figure}
\includegraphics[height=57mm]{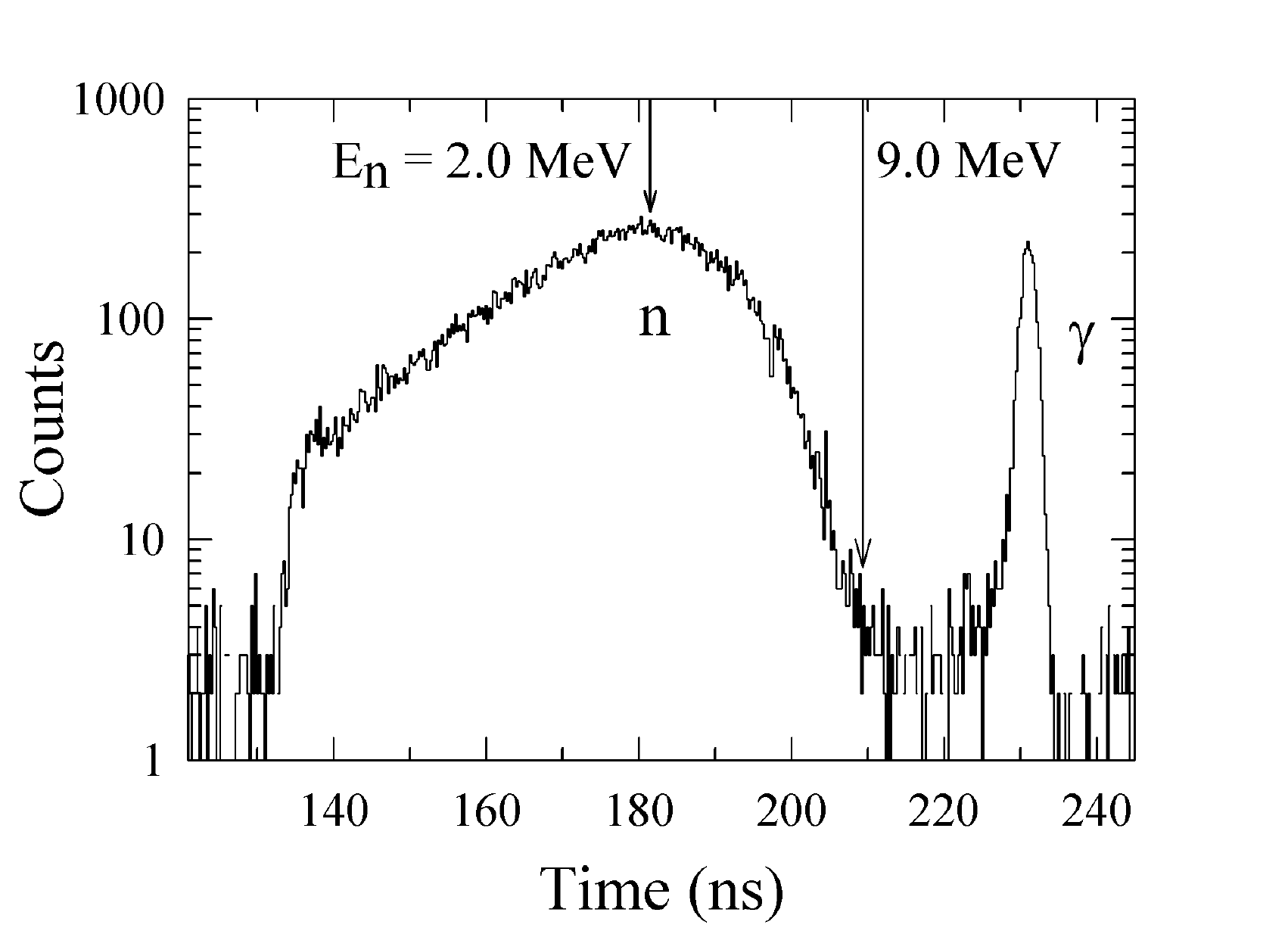}
\caption{Time of flight spectrum in $^7$Li + $^{205}$Tl reaction for the central energy bin
of alpha particles. The arrows indicate the positions for two representative neutron energies.}
\label{fig2}
\end{figure}

\begin{figure}
\includegraphics[width=85mm,trim=0mm 0mm 0mm 0mm,clip]{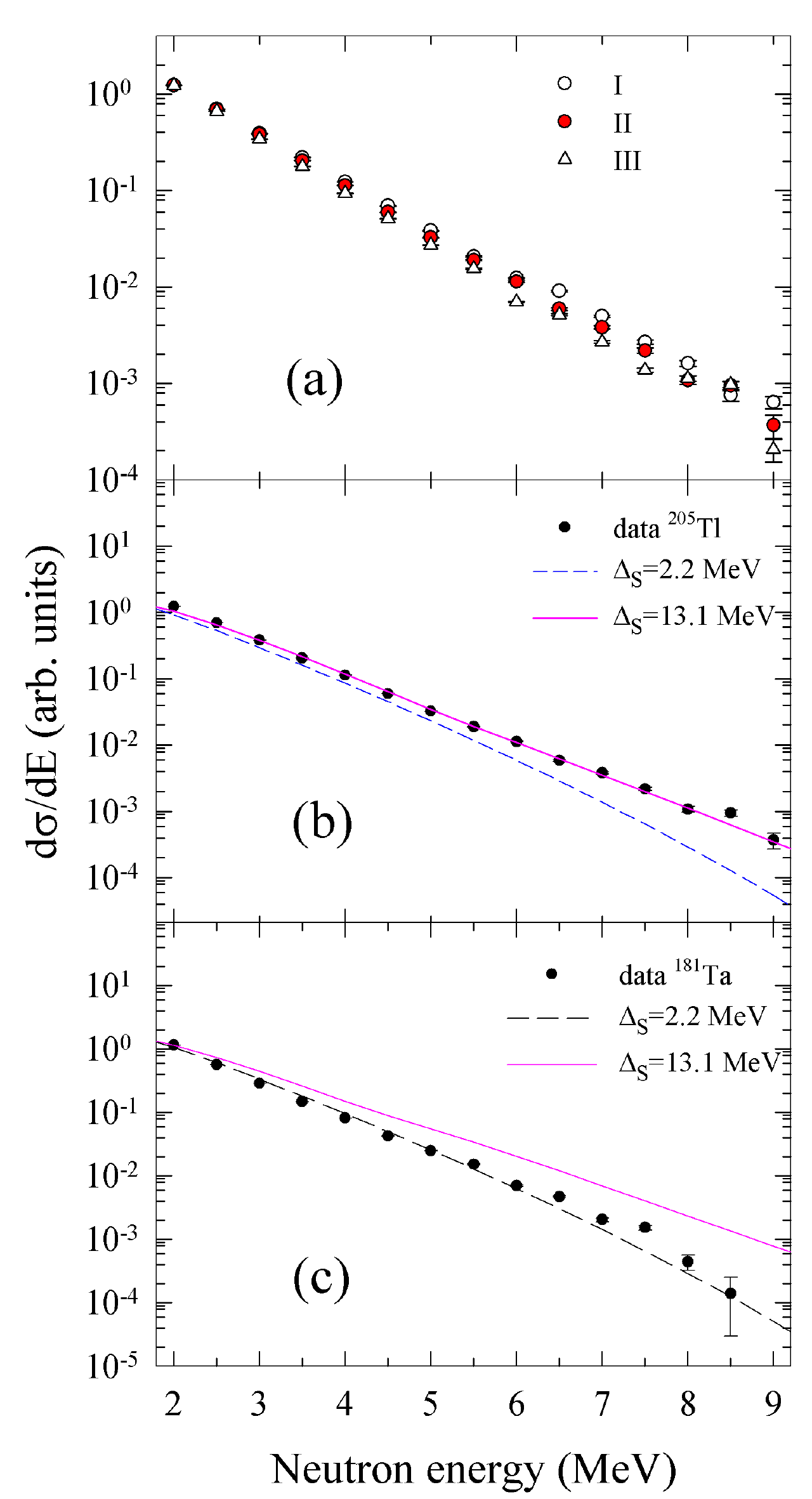}
\caption{(Color online)~(a)~Measured neutron spectra from $^{205}$Tl target for
three alpha bins I, II, III (see Fig.~\ref{fig1}), (b) and (c) show
measured neutron spectra from $^{205}$Tl and $^{181}$Ta targets and
statistical model (SM) calculations for the central alpha energy bin which 
corresponds to an excitation energy $\sim$ 21 MeV.}
\label{fig3}
\end{figure}

 Neutrons were detected using an array of 15 plastic detectors each of dimension
6~cm $\times$ 6~cm $\times$ 100~cm viewed by two photomultipliers (PMT), one at each end 
\cite{pcr}. The array was placed at a mean angle of 90$^{\circ}$ to the beam direction
and  at a distance of 1~m from the target. The neutron energy was measured using 
the time of flight (TOF) technique. The data were collected in an event by event mode
using a CAMAC based data acquisition system. The parameters recorded were (a)~left and right timing
of each plastic detector with respect to RF from the beam pulsing system using time to digital converters, (b)~integrated
charge of anode pulses (which is related to the electron equivalent energy, E$_{ee}$, deposited in the 
plastic detector) from the left and right PMTs using charge to digital converters, (c)~timing of CsI(Tl)
detectors with respect to the pulsed beam, (d)~energy  deposited in 
the CsI(Tl) detectors ($E_{CsI}$) and (e)~ZCT of the CsI detectors.

A typical ZCT~-~$E_{CsI}$ 2D-spectrum is shown in Fig.~\ref{fig1} displaying a clean separation of various groups of particles. 
The energy calibration of the CsI(Tl) detectors, in the range E$_{\alpha}\sim$ 5~-~25~MeV, was done 
using a $^{229}$Th alpha source and the $^{12}$C($^{12}$C,~$\alpha$)$^{20}$Ne 
reaction at E($^{12}$C)~=~24~MeV populating discrete states in $^{20}$Ne. For the latter measurement,
the carbon targets, backed by 1~-~3 mil thick Ta foils, were placed 23~cm upstream of the
centre of the reaction chamber and the detectors brought to 0$^\circ$ to reduce the kinematic energy
spread. The projected alpha energy spectrum for the $^{205}$Tl target is shown 
in the inset of Fig.~\ref{fig1}.
\begin{figure}
\includegraphics[width=85mm,trim=0mm 0mm 0mm 0mm,clip]{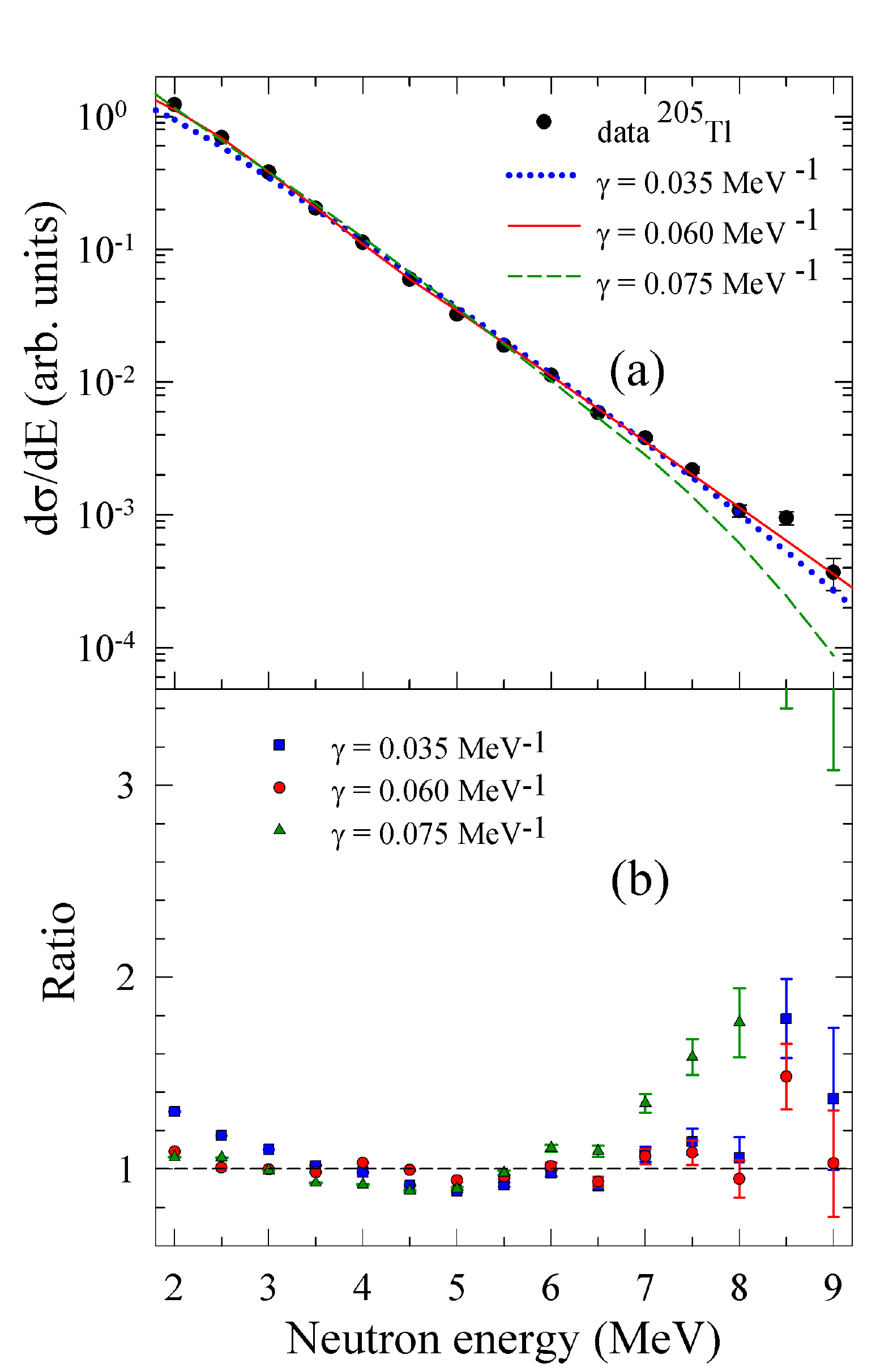}
\caption{(Color online) (a)~Comparison of data with SM calculation using $\Delta_S$~=~13.1~MeV (for $^{207}$Pb) and 11.7 MeV (for $^{206}$Pb) for three values of  $\gamma$ and (b)~ Ratio plot of data to fits for these $\gamma$  values.
} 
\label{fig4}
\end{figure}
The calibration of the energy deposited in the plastic detector (in E$_{ee}$) was 
done using Compton tagged recoil
electrons from  $^{137}$Cs and  $^{60}$Co $\gamma$-ray sources.
The time calibration was done using a precision time calibrator. 
The TOF, position information and geometric mean of the energy deposited for the neutron events in 
the plastic detector have been derived as in Ref.~\cite{pcr}.
In order to minimize the contribution of scattered neutrons a TOF dependent energy threshold 
(increasing with decreasing TOF) was used to obtain the final TOF spectra after subtracting the contribution
from the random coincidences.
A typical TOF spectrum is shown in Fig.~\ref{fig2}. 
The efficiency of the plastic detector as a function of incident neutron energy and 
energy threshold was calculated using a Monte Carlo simulation code \cite{pcr}. 
The efficiency corrected energy spectra of neutrons were 
derived from the TOF data.

The neutron energy spectra for the Tl target are shown in Fig.~\ref{fig3}(a) for three alpha energy bins, defined 
in Fig.~\ref{fig1}. An overall decrease in the slope of the spectra with the increase in alpha energy
(implying a decrease of E$_X$ in $^{208}$Pb) is consistent with the statistical nature of the neutron decay
from an equilibriated nucleus.
Fig.~\ref{fig3}(b) and (c) show the energy spectra for both the targets gated by the central
alpha energy bin. 

The statistical model~(SM) analysis of the spectra was done using the code CASCADE \cite{puhl} with 
the E$_X$ and J dependent NLD,
\[\rho(E_X,J)=\frac{2J+1}{12\sqrt{a}U^2}\left(\frac{\hbar^2}{2\Im}\right)^{3/2}e^{2\sqrt{aU}},\]
where $U = E_X - E_{\rm{rot}} - \Delta_P$, $\Delta_P$ is the 
pairing energy and $E_{\rm{rot}} = \left(\frac{\hbar^2}{2\Im}\right) J(J+1), \Im$ being the moment of inertia. 
The excitation energy dependence of the NLD parameter $a$, which includes the shell effect and its damping,
has been parameterised by Ignatyuk \cite{avi} as
\[a=\tilde{a}[1-\frac{\Delta_S}{U}(1 - e^{-\gamma U})].\] Here $\tilde{a}$ is the asymptotic
value of the NLD parameter in the liquid drop region, $\Delta_S$  is the shell 
correction energy, which is the difference between the experimental binding energy and that calculated from the LDM
and $\gamma$ is the damping parameter. 
Figs.~\ref{fig3}(b) and (c) show the calculated spectra using  $\tilde{a}$~=~A/8.5 MeV$^{-1}$ and
$\gamma$~=~0.055~MeV$^{-1}$~\cite{sch}. It is seen from the figure that 
a shell correction energy $\Delta_S$~=~13.1~MeV (for $^{207}$Pb ) fits the shape of neutron spectrum for the Tl target 
while $\Delta_S$~=~ 2.2~MeV does not. An opposite behaviour is seen 
for the Ta target. These values agree with those obtained from the experimental nuclear masses and the calculated LDM values\cite{myers}. The present data, therefore, is consistent with the shell correction energies derived from the nuclear masses. 

We now address the extraction of damping parameter from the present data. It may be pointed out that constraining all three parameters, $\tilde{a}$, $\Delta_S$ and $\gamma$, is not possible from the 
data addressing even a much wider excitation energy range. By fixing any two parameters the third one can be constrained. Since  the  shell correction energy is known with a reasonably good accuracy (within a few hundred keV \cite{myers}), we have fixed $\Delta_S$ and searched for an acceptable range of $\tilde{a}$ and $\gamma$. The shell correction energy was taken as 13.1 and 11.7 MeV for $^{207}$Pb and $^{206}$Pb, respectively. These two nuclei are only
relevant in the present case because the first two steps of neutron emission describe the full spectra. 

\begin{figure}
\includegraphics[width=75mm,height=60mm]{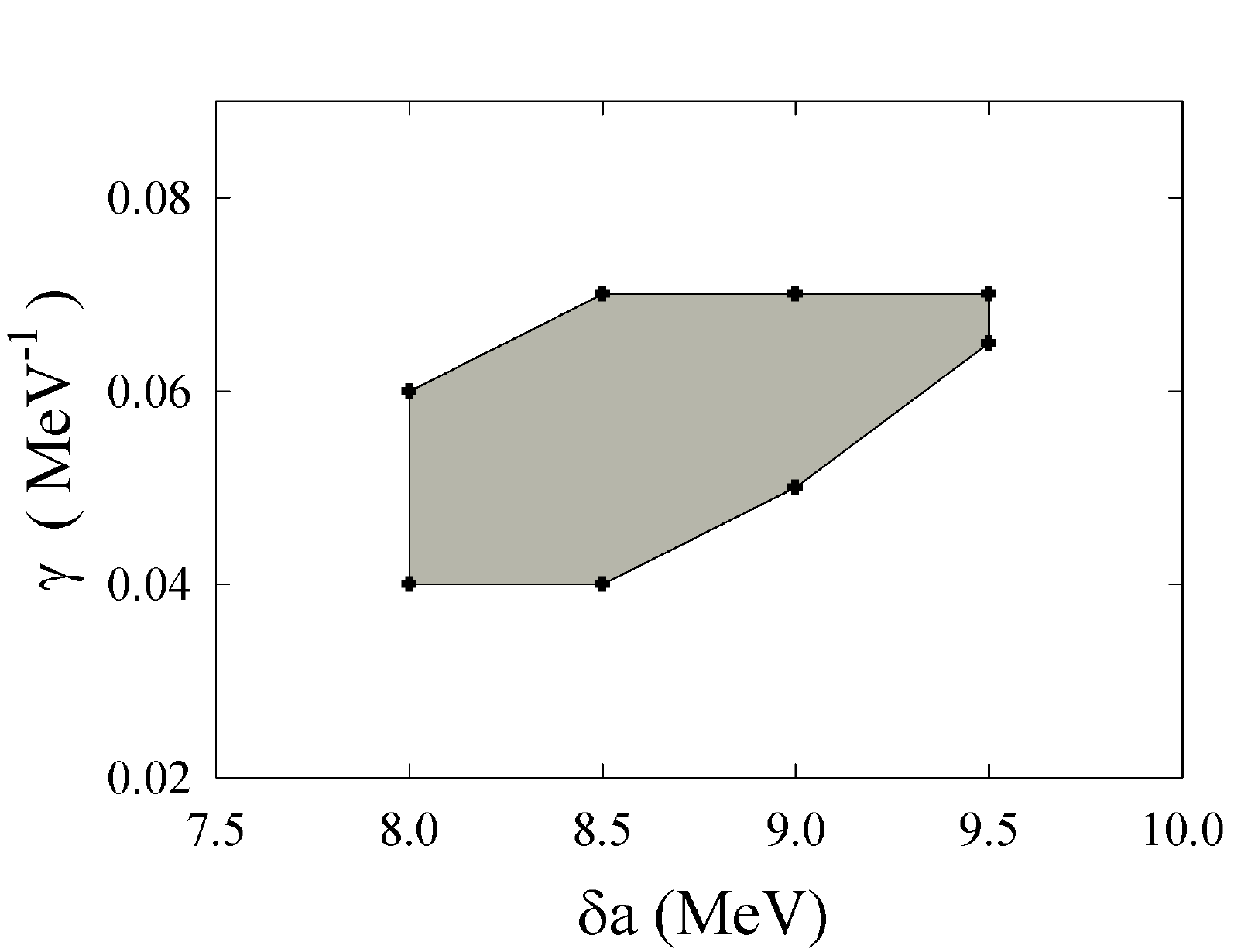}
\caption{Exclusion plot of $\delta a$~-~$\gamma$, where $\delta a$~=~A/$\tilde{a}$, for the shell correction energies quoted in Fig.~\ref{fig4}. The acceptable values are within the contour.}
\label{fig5}
\end{figure}

The calculations were performed with 
$\delta a$~(=~A/$\tilde{a}$) and $\gamma$ ranging from 6.5~-~11.0~MeV and 0.02~-~0.08 MeV$^{-1}$, respectively. 
Fig.~\ref{fig4}(a) shows statistical model fits for the central alpha energy bin for $\delta a$~=~8.5~MeV and three $\gamma$ values. The quality of the fits can also be judged from the ratio plots shown in Fig.~\ref{fig4}(b). Whereas a value of $\gamma$=0.060 MeV$^{-1}$ gives a good fit, the other two values can be discarded. It may be mentioned that
a change in shell correction energies up to 0.5~MeV has $<$2\% effect on the shape of the spectra. Similar analysis has been done for other two alpha energy bins. 
Fig.~\ref{fig5} shows a $\delta a$~-~$\gamma$ two dimensional exclusion plot, the region
inside the contour representing the acceptable range of parameter values for fitting the
present data. The criterion of rejection is based on both the relative $\chi^2$ values and the visual inspection of the fits over a range of E$_{\rm n}$=2~-~9~MeV. It can be seen from the figure that the acceptable range of $\delta a$ lies between 8.0 and 9.5~MeV.
The parameter $\gamma$ controlling the damping of the shell effect 
can be constrained to (0.060$^{+.010}_{-.020}$)~MeV$^{-1}$. This is different from the value extracted from the neutron 
resonance data $\it{viz.}$ (0.079~$\pm$~0.007)~MeV$^{-1}$~\cite{mughab}. This could be due to the differences in the angular momentum states sampled in the two works. Moreover, the present work addresses a specific nuclear region whereas
the analysis of Ref.~\cite{mughab} is global in character.
 
Finally we discuss the possible sources of uncertainties in the present experimental method. While the major contribution to the $\alpha$-coincident neutron spectra is expected to 
arise from triton transfer-fusion reaction, there are other direct processes that could contribute. The 
proton pickup and 2-neutron transfer cross sections are small~\cite{maha2} and can be ignored. 
A Monte Carlo calculation of the alpha-neutron coincidence spectrum reveals that the
contribution from the one neutron and one proton transfer is a small fraction ($<$ 5\%) in the 
region of interest, even if the cross sections are the same as that of the main reaction.
The most relevant reaction is the  deuteron transfer followed by $^5$He breakup. However, the spectroscopic factor for the d+$^5$He configuration is expected to be much smaller than the t+$^4$He 
configuration~\cite{wir} leading to a small contribution from this reaction.   

In conclusion,  we have for the first time measured the effect of the shell correction 
on the level density parameter over a range of excitation energy where the effect of 
damping is significant. The experimental results  show that the shell 
correction is indeed necessary to explain the data and is pronounced in the Pb region. 
The shell damping factor $\gamma$~=~(0.060$^{+.010}_{-.020}$)~MeV$^{-1}$ has been extracted from the data.
A precise measurement of the damping parameter in heavy magic nuclei will be an useful input in the 
current research on the formation of super heavy nuclei from heavy ion fusion reactions.
The precision of the present method can be improved by using a sharper time profile of 
the pulsed beam, pulse shape discrimination based neutron detectors
and a larger angular coverage. 

We thank S. S. Kapoor for his valuable suggestions. 
We acknowledge the help of the target laboratory staff, the PLF staff for smooth operation of the accelerator and 
Nitali Dash, R. Kujur and M. Pose for their support during the experiment.

\end{document}